# The impact of the substrate on the opto-thermal response of thin metallic targets following irradiation with femtosecond laser pulses

G. D. Tsibidis [1,2♣], and E. Stratakis [1,3♦]

[1] *Institute of Electronic Structure and Laser (IESL), Foundation for Research and Technology (FORTH), N. Plastira 100, Vassilika Vouton, 70013, Heraklion, Crete, Greece*

[2] *Department of Materials Science and Technology, University of Crete, 71003 Heraklion, Greece*

[3] *Department of Physics, University of Crete, 71003 Heraklion, Greece*

Email: [♣]tsibidis@iesl.forth.gr;  [♦]stratak@iesl.forth.gr

## Abstract

Femtosecond pulsed lasers have been widely used over the past decades due to their capability to fabricate precise patterns at the micro- and nano- lengths scales. A key issue for efficient material processing is the determination of the laser parameters used in the experimental set ups. Despite a systematic investigation that has been performed to highlight the impact of every parameter independently, little attention has been drawn on the role of the substrate material on which the irradiated solid is placed. In this work, the influence of the substrate is emphasised for films of various thicknesses which demonstrates that both the optical and thermophysical properties of the substrate affect the thermal fingerprint on the irradiated film while the impact is manifested to be higher at smaller film sizes. Two representative materials, silicon and fused silica have been selected as typical substrates for thin films of different optical and thermophysical behaviour (gold and nickel) and the thermal response and damage thresholds are evaluated for the irradiated solids. The pronounced influence of the substrate is aimed to pave the way for new and more optimised designs of laser-based fabrication set ups and processing schemes.

## I. INTRODUCTION

The employment of femtosecond (fs) pulsed laser sources for material processing has received considerable attention due to the important technological applications [1-7]. A systematic analysis of the ultrafast phenomena that occur following excitation of solids with fs pulses and resulting thermal effects are important to facilitate the control of laser energy towards fabricating application-based topographies. Thus, a thorough understanding of the underlying complex physical mechanisms appears to be significant both from an application and fundamental point of view; therefore, consistent methodologies have been developed over the past decades to explore in detail the multiscale phenomena [8-23].

A crucial issue for efficient material processing is the accurate determination of the damage threshold [24-27]. This parameter is frequently associated with the laser peak fluence at which minimal damage occurs on the surface of the irradiated solid. Various standard methodologies for an accurate estimation of the damage threshold have been developed and presented in previous works [24-27]. From a theoretical point of view, the prediction of the damage threshold was performed considering a thermal criterion (i.e. identification of conditions that lead to ablation or melting) [8, 28-30] through the use of the classical Two Temperature Model (TTM) [31] which describes the electron-phonon temperature dynamics and relaxation process [32]. Alternative studies also included the employment of atomistic continuum models with the combination of Molecular Dynamics and TTM [20, 32].

Nevertheless, most of the current research has been focused on the investigation of damage conditions on bulk materials. On the other hand, there has been an increasing interest in patterning of thin solid films (of sizes comparable to the optical penetration depth) for various applications related to optics, healthcare, sensing, environment, energy [33-43], and therefore a detailed exploration of the ultrafast dynamics and damage threshold evaluation for such materials is required. Although results have been reported for a variety of materials [44-48], it has been shown that the optical properties of thin materials deviate from those of bulk solids as the thickness decreases to sizes comparable to the optical penetration depth [47]. Thus, the amount of the absorbed energy will also differ from whether the material is considered as a bulk solid which, in turn, is expected to be reflected on the damage threshold. In previous reports, it was shown that thinner films appear to inhibit the electron diffusion which delays the electron-phonon coupling and leads to lower damage thresholds [44, 46, 47, 49, 50].

Another parameter that is usually ignored during the investigation of the ultrafast dynamics in the evaluation of the thermal response of double- (or multi-) layered materials is the impact of the substrate. It has been demonstrated that in multi-layered films, the surface temperature (and eventually, the damage threshold) is influenced from the properties of the constituent materials [51-56]. Nevertheless, to the best of our knowledge, no previous study has been reported on how the opto-thermal characteristics of a substrate will affect the damage threshold of the irradiated material. Whether thermal effects following irradiation of thin films (of size close to the optical penetration depth) should also involve the contribution from the opto-thermal features of the substrate requires a thorough investigation. Furthermore, it has been shown that a generic conclusion for the thermal response of a thin film and correlation of the damage threshold with thickness are not possible but they are material dependent [50]. Therefore, a threshold at which the substrate does not influence any longer the damage threshold of the upper layer is not expected to be identical to all materials.

The elucidation of the aforementioned issues is of paramount importance not only to understand the complex physical mechanisms of laser-matter interactions and ultrafast electron dynamics for thin films but also to associate the resulting thermal effects with targeted patterning strategies. To this end, in this work, we present a theoretical model that can be used to describe the ultrafast dynamics, thermal effects and damage thresholds for Au and Ni of increasing thickness and different substrates (Si and SiO$_2$) (Section II). The evaluation of variation of the dielectric parameters as a function of the material thickness is performed through the application of a 'multiple reflection' algorithm [57]. Relaxation processes are described through the employment of a TTM and a melting-point-based thermal criterion is used to determine the damage threshold. A systematic analysis of the results is discussed in Section III while concluding remarks follow in Section IV.

## II. THEORETICAL MODEL

To describe the ultrafast dynamics for an irradiated material following exposure to fs pulses, a theoretical framework is employed to provide a detailed investigation of the excitation and thermal response of a double-layered structure (thin film/SiO$_2$) and (thin film/Si). The traditional theoretical model used to describe the above process is based on the use of a Two Temperature Model (TTM) [31]. In this work, for the sake of simplicity, an 1D-TTM is employed to describe the thermal effects due to heating of the thin films with laser pulses of wavelength $\lambda_L$=1026 nm and pulse duration equal to $\tau_p$=170 fs [54]

$$
\begin{aligned}
C_e^{(m)} \frac{\partial T_e^{(m)}}{\partial t} &= \frac{\partial}{\partial z}\left(k_e^{(m)} \frac{\partial T_e^{(m)}}{\partial z}\right) - G_{eL}^{(m)}\left(T_e^{(m)} - T_L^{(m)}\right) + S^{(m)} \\
C_L^{(m)} \frac{\partial T_L^{(m)}}{\partial t} &= \frac{\partial}{\partial z}\left(k_L^{(m)} \frac{\partial T_L^{(m)}}{\partial z}\right) + G_{eL}^{(m)}\left(T_e^{(m)} - T_L^{(m)}\right) \\
C_e^{(S)} \frac{\partial T_e^{(S)}}{\partial t} &= \frac{\partial}{\partial z}\left(k_e^{(S)} \frac{\partial T_e^{(S)}}{\partial z}\right) - G_{eL}^{(S)}\left(T_e^{(S)} - T_L^{(S)}\right) + S^{(S)} \\
C_L^{(S)} \frac{\partial T_L^{(S)}}{\partial t} &= \frac{\partial}{\partial z}\left(k_L^{(S)} \frac{\partial T_L^{(S)}}{\partial z}\right) + G_{eL}^{(S)}\left(T_e^{(S)} - T_L^{(S)}\right)
\end{aligned}
\tag{1}
$$

where the subscript '$m$' (or '$S$') indicates the thin film (or substrate). In Eqs.1, $T_e^{(m)}$ and $T_L^{(m)}$ stand for the electron and lattice temperatures, respectively, of the metallic thin film. The thermophysical properties of the metal such as the electron $C_e^{(m)}$ (or lattice $C_L^{(m)}$) volumetric heat capacities, electron $k_e^{(m)} \left(= k_{e0}^{(m)} \frac{B_e T_e^{(m)}}{A_e \left(T_e^{(m)}\right)^2 + B_e T_L^{(m)}}\right)$ (or lattice $k_L^{(m)}$=0.01$k_e^{(m)}$) heat conductivities, the electron-phonon coupling strengths $G_{eL}^{(m)}$, $A_e$, $B_e$ and other model parameters that show in the first two equations of Eqs.1 are listed in Table 1.

On the other hand, the quantity $S^{(m)}$ represents the source term that represents the energy that the laser source gives to the solid which is assumed to be sufficient to generate excited carriers on the thin film. The following processes are considered: (i) a portion of the energy is absorbed in the material while part of the laser energy is transmitted into the substrate, (ii) the reflectivity and transmissivity of the irradiated material are influenced by a *multiple reflection process* between the two interfaces (air/metal and metal/substrate),

Some special attention is required for the transmission of energy in the substrate. More specifically, in bulk metals (or in metals of thickness) the transmitted energy in the substrate is minimal as almost of all of the energy that is not reflected is absorbed from the metal; thus, a temperature rise of the substrate through excitation can be ignored as the transmitted energy into the substrate is not sufficiently high to generate excited carriers. Therefore, the third equation of Eqs.1 can be ignored while the fourth can be simplified by $C_L^{(S)} \frac{\partial T_L^{(S)}}{\partial t} = \frac{\partial}{\partial z}\left(k_L^{(S)} \frac{\partial T_L^{(S)}}{\partial z}\right)$ where $T_L^{(S)}$, $C_L^{(S)}$, $k_L^{(S)}$ stands for the substrate temperature, volumetric heat capacity and heat conductivity, respectively. By contrast, recent simulations have indicated that for thicknesses of the order of the penetration depth, a more rigorous analysis is required as the transmitted energy is not negligible [50]. To this end, the third and fourth equations in Eqs.1 are replaced with the corresponding TTM equations for semiconductors [10] and dielectrics [17].

It is noted that the source term $S^{(m)}$ which is used to excite a metallic surface of thickness $d$ is given from the following formula [47]

$$S^{(m)} = \frac{(1-R-T)\sqrt{4\log(2)}F}{\sqrt{\pi}\tau_p(\alpha^{-1}+L_b)} \exp\left(-4\log(2)\left(\frac{t-3t_p}{t_p}\right)^2\right) \frac{\exp(-z/(\alpha^{-1}+L_b))}{(1-\exp(-d/(\alpha^{-1}+L_b)))} \tag{2}$$

where $R$ and $T$ stand for the reflectivity and transmissivity, respectively. On the other hand, $L_b$ corresponds to the ballistic length, $\alpha$ is the absorption coefficient which is wavelength dependent, and $F$ is the peak fluence of the laser beam. The ballistic transport is also included in the expression as it has been demonstrated that it plays significant role in the response of the material [47]. Special attention is required for the ballistic length as in previous works, it has been reported that for bulk materials, $L_b$ in $s/p$-band metals is large ($L_b^{(Au)}$=100 nm [47]) while for the $d$-band metals such as Ni it is of the same order as their optical penetration depth [47].

The multiple reflection theory is used to calculate $R$ and $T$ and the absorbance $A=1-R-T$ [57]. Thus, the following expressions are employed to calculate the optical properties for a thin film on a substrate (for a $p$-polarised beam)

$$R = |r_{dl}|^2, \quad T = |t_{dl}|^2 Re(\widetilde{N}_S), \quad r_{dl} = \frac{r_{am}+r_{mS}e^{2\beta j}}{1+r_{am}+r_{mS}e^{2\beta j}}, \quad t_{dl} = \frac{t_{am}t_{mS}e^{\beta j}}{1+r_{am}+r_{mS}e^{2\beta j}}, \quad \beta = 2\pi d/\lambda_L \tag{3}$$

$$r_{CD} = \frac{\widetilde{N}_D - \widetilde{N}_C}{\widetilde{N}_D + \widetilde{N}_C}, \quad t_{CD} = \frac{2\widetilde{N}_C}{\widetilde{N}_D + \widetilde{N}_C} \tag{4}$$

where the indices $C=a,m$ and $D=m,S$ characterise each material ('$a$', '$m$', '$S$' stand for 'air', 'metal', 'substrate', respectively). The complex refractive indices of the materials such as air, metal and substrate are denoted with $\widetilde{N}_a = 1$, $\widetilde{N}_m = Re(\widetilde{N}_m) + Im(\widetilde{N}_m)j$, $\widetilde{N}_S = Re(\widetilde{N}_S) + Im(\widetilde{N}_S)j$, respectively. Given that fused silica glass or silicon are used as the substrate material, the complex refractive indices for the materials are considered in these simulations at $\lambda_L = 1026$ nm (i.e. $Re(\widetilde{N}_s) = 1.4501$ and $3.5632 + 0.00027806j$ for SiO$_2$ [17] and Si [58], respectively).

To obtain the dielectric function for each metal, the Drude-Lorentz model is used which is based on the analysis by Rakic *et* al. (where both interband and intraband transitions are assumed) [59]. As the optical parameters of an excited material is expected to vary during the excitation process [21], to introduce the transient change, a temporally varying expression of the dielectric function is provided by including a temperature dependence on the reciprocal of the electron relaxation time $\tau_e$ (i.e. $\tau_e = \left[A_e\left(T_e^{(m)}\right)^2 + B_e T_L^{(m)}\right]^{-1}$) [60]. The values of the refractive indices of the metals in this study (at 300 K are given in Table 1).

The volumetric heat capacity of fused silica glass is $C_L^{(S)}\big|_{SiO_2}$ =1.6×10$^6$ Jm$^{-3}$K$^{-1}$ while the heat conductivity of fused silica glass is equal to $k_L^{(S)}\big|_{SiO_2}$ =1.38 Wm$^{-1}$K$^{-1}$ [17]. Similarly, the thermophysical properties for Silicon are $C_L^{(S)}\big|_{Si} = 10^6 \times \big[1.978 + 3.54 \times 10^{-4} T_L^{(S)} - 3.68\left(T_L^{(S)}\right)^{-2}\big]$ Jm$^{-3}$K$^{-1}$ and $k_L^{(S)}\big|_{Si} = 1585 \times 10^2 \left(T_L^{(S)}\right)^{-1.23}$ Wm$^{-1}$K$^{-1}$ [61].

The set of equations Eqs.1-4 are solved by using an iterative Crank-Nicolson scheme based on a finite-difference method. It is assumed that the system is in thermal equilibrium at $t=0$ and, therefore, $T_e^{(m)}(z, t = 0) = T_L^{(m)}(z, t = 0) = 300$ K. A thick substrate is, also, considered (i.e. $k_L^{(S)} \frac{\partial T_L^{(S)}}{\partial z} = 0$) while adiabatic conditions are applied on the surface of the metallic surface (i.e. $k_e^{(m)} \frac{\partial T_e^{(m)}}{\partial z} = 0$). Finally, the following boundary conditions are assumed on the interface between the top layer and the substrate: $k_L^{(m)} \frac{\partial T_L^{(m)}}{\partial z} = k_L^{(S)} \frac{\partial T_L^{(S)}}{\partial z}, k_e^{(m)} \frac{\partial T_e^{(m)}}{\partial z} = k_e^{(S)} \frac{\partial T_e^{(S)}}{\partial z}, T_L^{(m)} = T_L^{(S)}$.

| Parameter | Material | |
|---|---|---|
| | Au | Ni |
| $\widetilde{N}_m$ | DL [59] | DL [59] |
| $G_{eL}^{(m)}$ [Wm$^{-3}$K$^{-1}$] | Ab-Initio [62] | Ab-Initio [62] |
| $C_e^{(m)}$ [Jm$^{-3}$K$^{-1}$] | Ab-Initio [62] | Ab-Initio [62] |
| $C_L^{(m)}$ [$\times 10^6$ Jm$^{-3}$K$^{-1}$] | 2.48 [47] | 4.3 [47] |
| $k_{e0}^{(m)}$ [Wm$^{-1}$K$^{-1}$] | 318 [47] | 90 [47] |
| $A_e$ [$\times 10^7$ s$^{-1}$K$^{-2}$] | 1.18 [52] | 0.59 [52] |
| $B_e$ [$\times 10^{11}$ s$^{-1}$K$^{-1}$] | 1.25 [52] | 1.4 [52] |
| $T_{melt}$ [K] | 1337 [52] | 1728 [52] |

Table 1: Optical and thermophysical properties of Au and Ni (*DL* stands for Drude-Lorentz model).

### III. RESULTS AND DISCUSSION

The ultrafast dynamics and the optical and thermal response of the irradiated material are described through the use of the theoretical model that was presented in Section II. The ultrafast dynamics of two different materials one s/p-band metal (Au) and a d-band metal (Ni) with single laser pulses of $\lambda_L = 1026$ nm was explored while similar conclusions can be deduced at different laser wavelengths and other materials.

As the predominant objective of the current work is to highlight the impact of the substrate on the ultrafast dynamics, a thorough investigation is performed to analyse both the optical and thermal response of the irradiated metals for Si and SiO$_2$ substrates. A reasonable argument that the influence of the substrate should be more pronounced if the heat affected region is closer to the substrate leads us to correlate the thickness of the irradiated solid with the type of the substrate. Our simulation results for two distinctly different thicknesses, 30 nm and 400 nm, illustrate remarkably contrasting behaviour of the electron and lattice temperatures for the two metals assuming different substrates (Figs.1-2). More specifically, the theoretical predictions demonstrate that for *d* of the size of the optical penetration depth there exists a significant influence of the role of the substrate (Fig.1a, Fig.2a). By contrast, for larger *d*, the impact of the substrate diminishes gradually and for *d*=400 nm (where the metal film can be identified as a rather 'bulk' material [50]), the ultrafast dynamics on the surface of the metal is independent of the substrate (Fig.1a, Fig.2a). It is noted that *F*=0.3 J/cm$^2$ and *F*=0.1 J/cm$^2$ were used for Au and Ni, respectively.

Before we attempt to explain why this discrepancy appears and why Si substrates favours larger lattice temperatures on the metals surface (Figs.1-2), we notice, firstly, a longer equilibration process between the electron and lattice temperatures at thinner samples; this behaviour appears to be independent of the substrate. As noticed in previous reports [47, 50] this is due to the fact that for smaller thicknesses, the electron diffusion is inhibited and therefore high-energetic electrons remain in the affected region. As a result, the electron-phonon scattering constitutes the dominant relaxation mechanism and the relaxation delay results from the need of multiple scattering events for equilibration of the energies of the electron and lattice subsystems. A comparison of the dynamics and the temperature evolution for both metals manifest a faster relaxation process for Ni than for Au and this monotonicity holds also for thinner films. This is due to the fact that the electron-phonon coupling for Ni is substantially larger at small temperatures [62] (in principle, $G_{eL}^{(m)}$ exhibits an opposite monotonicity at increasing electron temperature compared to Au) which accelerates the relaxation process. Furthermore, theoretical simulations show that the $SiO_2$ substrate delays further the equilibration procedure.

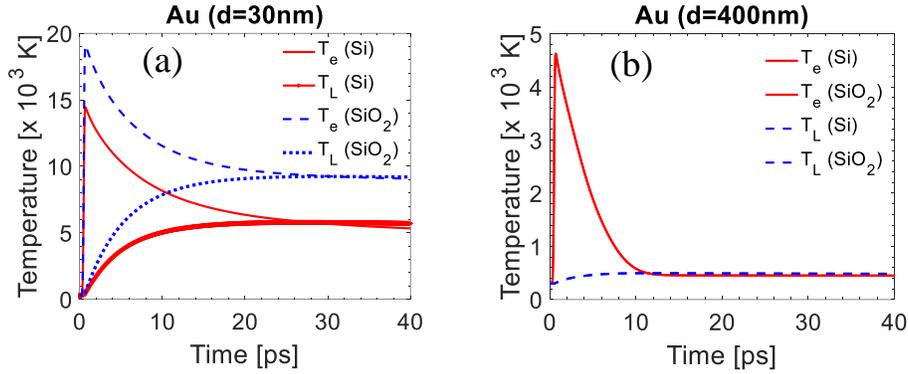

Figure 1: Electron and lattice temperature evolution for Au of thickness (a) $d$=30 nm, (b) $d$=400 nm) for Si and $SiO_2$ substrates ($\lambda_L$=1026 nm, $F$=0.3 J/cm$^2$).

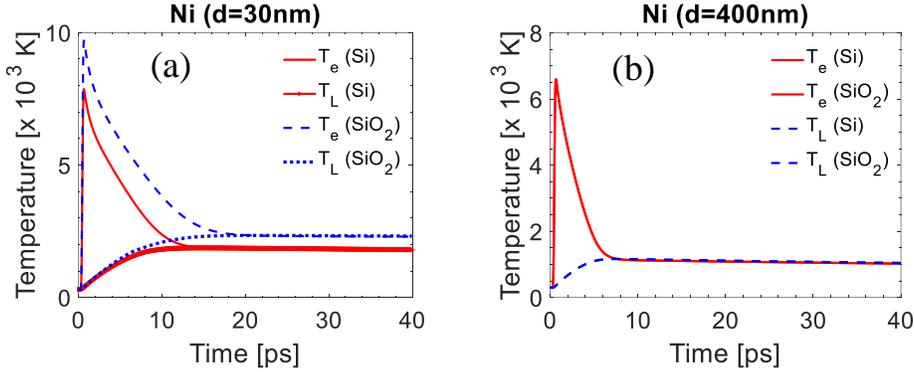

Figure 2: Electron and lattice temperature evolution for Ni of thickness (a) $d$=30 nm, (b) $d$=400 nm) for Si and $SiO_2$ substrates ($\lambda_L$=1026 nm, $F$=0.1 J/cm$^2$).

One important question that rises is whether the thermophysical properties of the substrate and/or the difference the heat conductivities and capacities of the two-layered material (i.e. metal/substrates) when Si or $SiO_2$ is the substrate accounts for the distinct behaviour illustrated in Figs.1-2. It is noted, firstly, that a significant difference between the two substrates is that Si is characterised by a larger lattice heat conductivity which implies that a substantially larger energy generated on the metal diffuses to the Silicon lattice that could lead to smaller maximum lattice temperatures on the surface of the metal. Nevertheless, this plausible explanation should probably explain the thermal behaviour at larger time points; by contrast, even after the end of the pulse, both the generated temperatures of the excited electrons ($T_e$) and lattice ($T_L$) are higher for Si substrate. The larger maximum $T_e$ for Si substrate demonstrates that a higher excitation level is reached which indicates that the optical response of the irradiated metals differs when Si or $SiO_2$ are used as the solid substrates; this assumption appears, also, to be justified given that the discrepancy disappears for large thicknesses.

To account for potential different amount of energy absorption due to a change in the optical response in the two cases, we have performed a thorough analysis of the optical parameters of the complex (metal/substrate), the absorbance and reflectivity. In a recent report, it has been shown that due to a multiple reflection process, these optical parameters for noble (such as Au) and transition (such as Ni) metals differ substantially for thin films of sizes of the order of the optical penetration depth or smaller [50].

Simulations results indicate that there is distinct increase of the reflectivity for Au and Ni as the thickness size becomes larger and it saturates to a value that characterizes bulk materials. The theoretical model and the employment of Eqs.3-4 predict substantially smaller reflectivity values for thicknesses close to the penetration depth (~15-20 nm) compared to the values for thicker (or bulk) materials. These results demonstrate a significant influence of the multiple reflection process inside the metal on the optical properties at various thicknesses (Figs.3a-b). Similar conclusions can be derived for the transmissivity and therefore the absorbance of the metal

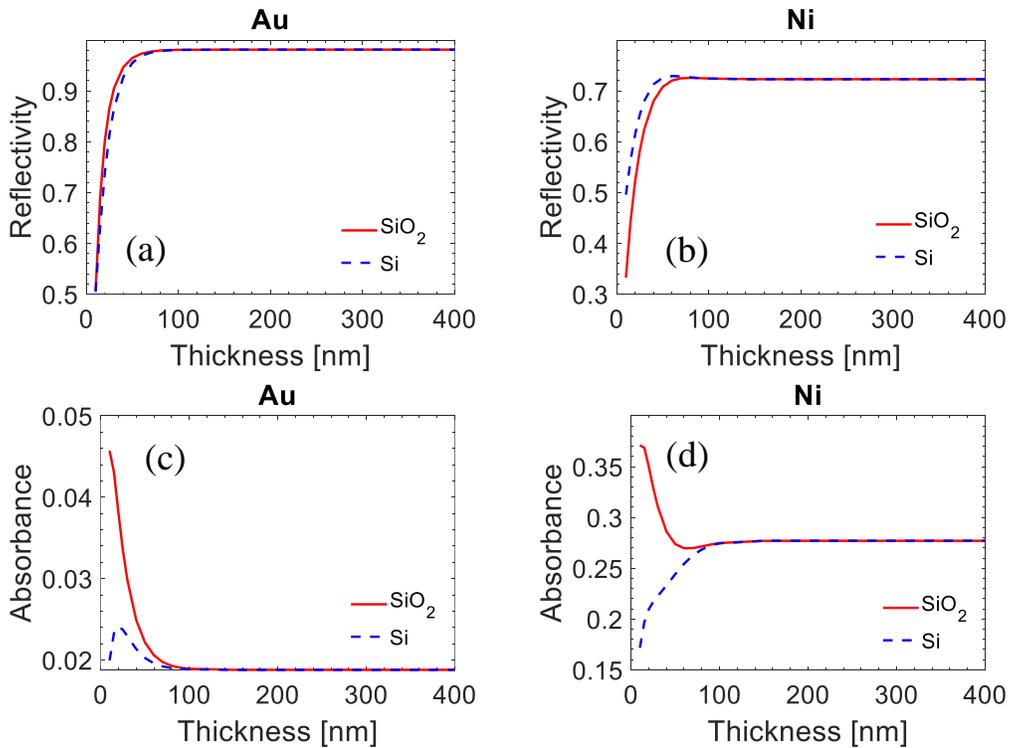

Figure 3: Reflectivity and absorbance for Au and Ni for Si/$SiO_2$ substrates ($\lambda_L$=1026 nm).

(Figs.3c-d). To avoid any confusion, it is noted that the values of the optical parameters illustrated in Fig.3 have been evaluated at 300 K, however, the distinct behavior can provide a conclusive picture of the impact of the substrate and the thickness of the metal film. On the other hand, and in an attempt to reveal the role of the substrate, results for Au and Ni (Fig.3a-b) show that the Si substrate leads to a larger reflectivity compared to the $SiO_2$ for Nickel (up to 50% at small thicknesses). By contrast, the induced reflectivity of Au for Si substrate is smaller (up to 10%). This discrepancy disappears as the thickness of the metal film increases. Furthermore, for both materials, a larger absorbance of the irradiated metal occurs if $SiO_2$ substrate is used that implies that a smaller laser energy amount is finally absorbed from the metal (Fig.3c-d). As the metal thickness increases and reaches the size of a bulk material, the laser pulse does not 'see' the substrate and the optical parameters are independently of the substrate. It is noted that the simulation results and nonnegligible transmittance coefficient (1-*R*-*A*) (Fig.3) for small metal thicknesses manifest the need to include appropriate revised versions of the third and fourth equations in Eqs.1 [10,17] as described in Section II.

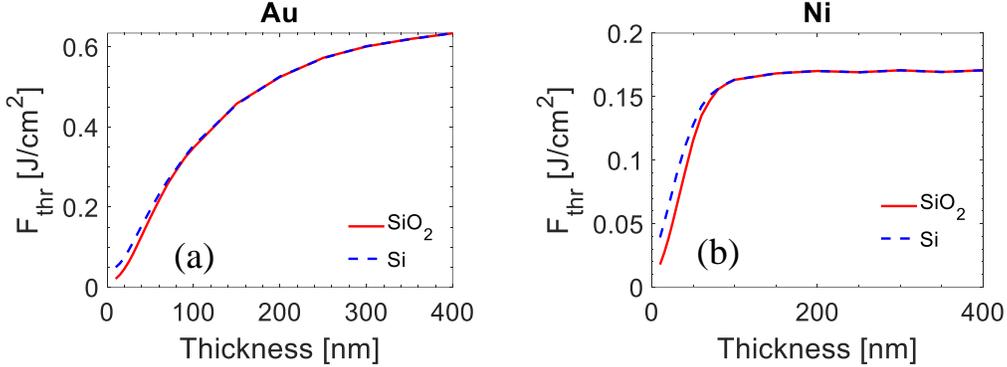

Figure 4: Damage threshold for various thicknesses for Au and Ni for Si/$SiO_2$ substrates ($\lambda_L$=1026 nm).

Given the remarkable difference in the absorbed energy at small thicknesses in the presence of Si or $SiO_2$ substrates, the question is whether the impact of the substrate is also projected on the thermal response of the irradiated material. The use of the multiscale physical model presented in Section II could be employed to correlate the induced thermal effects with the onset of the material damage. In this work, a melting-point-based thermal criterion is considered to determine whether material is damaged. More specifically, the minimum laser fluence, $F_{thr}$, which is sufficient to melt the material is chosen as the damage threshold. Simulations (Fig.4) demonstrate an increase of $F_{thr}$ with increasing thickness for all materials which is due to the decrease of the produced lattice temperature on the surface of the material due to, primarily, electron diffusion and ballistic transport. This agrees with the discussion in this section and the fact that electron diffusion is facilitated if the metal thickness increases. On the other hand, results show that there is a threshold value for the thickness $d_{thr}$ after which the material behaves more as a bulk solid, a saturation is reached and, therefore, further increase of $d_{thr}$ does not influence $F_{thr}$. Thus, for larger thicknesses, $F_{thr}$ exhibits an asymptotic behaviour close to the damage threshold of the bulk material. Experimental data in previous reports confirm also this behaviour [44, 46, 48]. To compare the influence of the substrate, results in Fig.4 illustrate the increase of the damage threshold for small thicknesses for a Si substrate. This is explained by the smaller energy which is absorbed from the material compared to a $SiO_2$ substrate. As expected, this discrepancy disappears for larger thicknesses. Simulations results that for both materials a rise of the damage threshold can be achieved by 50% if a Si substrate is used. Results also show that for Ni, a thickness equal to ~100 nm constitutes a saturation value above which the substrate does not play significant role. By contrast, the saturation value for Au is approximately equal to ~80 nm. Certainly, a generic conclusion for the saturation value is not possible for any material.

To implement the multiscale model and provide accurate estimates of the damage threshold in various conditions, it is important to evaluate precisely the absorbed energy and optical parameters. It is known that a consistent methodology for an accurate description of the ultrafast dynamics requires a transient calculation of the dielectric function. As pointed out in the description of the theoretical framework, a Drude-Lorentz model was used to obtain the dielectric function for each metal based on the approach by Rakic *et al*. (where both inter-band and intra-band transitions are assumed) [59]. In that analysis, the oscillator lengths and spectral widths for the Lorentzian terms that are used were obtained through fitting with experimental data and it was assumed that they do not vary during excitation conditions and at higher electron temperatures. On the other hand, the transient character of the dielectric function was introduced through the inclusion of the relaxation time $\left[A_e \left(T_e^{(m)}\right)^2 + B_e T_L^{(m)}\right]^{-1}$ [60]. Certainly, this approach provides a dynamic behaviour into the optical properties of the material and results in various reports have shown that the approximation can lead to a satisfactory agreement with experimental observations in some conditions [54, 63]. Nevertheless, a more precise investigation would require the employment of more rigorous approaches that reveal not only a time dependent variation but also an electron temperature dependence of the dielectric properties. Such approaches have been developed that were based on the use of first principles and Density Functional Theories to describe the ultrafast dynamics of various materials ranging from metals [21] to semiconductors [22] and they could be incorporated into a future and more comprehensive revised model.

## IV. CONCLUSIONS

The emphasis of the current work was on the role of the substrate of a metallic solid that is irradiated with fs laser pulses. Theoretical results demonstrated the impact of the thickness both on the optical properties and the damage threshold and they revealed a distinct impact of the substrate. More specifically, for two widely used metals in various applications (Au and Ni), simulations revealed an enhanced energy absorption from thin films if fused silica substrates are used compared to results for Silicon substrates. The theoretical predictions showed that larger excitation levels and produced lattice temperatures for fused silica substrates are capable to lead to smaller damage threshold values as big as 50% larger for both materials while the discrepancy disappears at substantially thicker films. The presented model is aimed to provide a tool for an accurate determination of the damage threshold of metals, which is important for a plethora of laser manufacturing approaches. The demonstrated influence of the substrate is aimed to pave the way for new and more optimised designs of laser-based fabrication set ups and processing schemes.

**ACKNOWLEDGEMENTS**

The authors acknowledge support by the European Union's Horizon 2020 research and innovation program through the project *BioCombs4Nanofibres* (grant agreement No. 862016). G.D.T and E.S. acknowledge funding from *HELLAS-CH* project (MIS 5002735), implemented under the "Action for Strengthening Research and Innovation Infrastructures" funded by the Operational Programme "Competitiveness, Entrepreneurship and Innovation" and co-financed by Greece and the EU (European Regional Development Fund) while G.D.T acknowledges financial support from COST Action *TUMIEE* (supported by COST-European Cooperation in Science and Technology).

**REFERENCES**

[1] E. Stratakis *et al.*, Materials Science and Engineering: R: Reports **141**, 100562 (2020).
[2] A. Y. Vorobyev and C. Guo, Laser & Photonics Reviews **7**, 385 (2012).

[3] V. Zorba, E. Stratakis, M. Barberoglou, E. Spanakis, P. Tzanetakis, S. H. Anastasiadis, and C. Fotakis, Advanced Materials **20**, 4049 (2008).
[4] V. Zorba, L. Persano, D. Pisignano, A. Athanassiou, E. Stratakis, R. Cingolani, P. Tzanetakis, and C. Fotakis, Nanotechnology **17**, 3234 (2006).
[5] J.-C. Diels and W. Rudolph, *Ultrashort laser pulse phenomena : fundamentals, techniques, and applications on a femtosecond time scale* (Elsevier / Academic Press, Amsterdam ; Boston, 2006), 2nd edn.
[6] E. L. Papadopoulou, A. Samara, M. Barberoglou, A. Manousaki, S. N. Pagakis, E. Anastasiadou, C. Fotakis, and E. Stratakis, Tissue Eng Part C-Me **16**, 497 (2010).
[7] A. Papadopoulos, E. Skoulas, A. Mimidis, G. Perrakis, G. Kenanakis, G. D. Tsibidis, and S. E., Advanced Materials **31**, 1901123 (2019).
[8] G. D. Tsibidis, M. Barberoglou, P. A. Loukakos, E. Stratakis, and C. Fotakis, Physical Review B **86**, 115316 (2012).
[9] G. D. Tsibidis, C. Fotakis, and E. Stratakis, Physical Review B **92**, 041405(R) (2015).
[10] G. D. Tsibidis, E. Stratakis, P. A. Loukakos, and C. Fotakis, Applied Physics A **114**, 57 (2014).
[11] T. J. Y. Derrien, T. E. Itina, R. Torres, T. Sarnet, and M. Sentis, Journal of Applied Physics **114**, 083104 (2013).
[12] J. Z. P. Skolski, G. R. B. E. Romer, J. V. Obona, V. Ocelik, A. J. H. in 't Veld, and J. T. M. De Hosson, Physical Review B **85**, 075320 (2012).
[13] J. Bonse, J. Krüger, S. Höhm, and A. Rosenfeld, Journal of Laser Applications **24**, 042006 (2012).
[14] F. Garrelie, J. P. Colombier, F. Pigeon, S. Tonchev, N. Faure, M. Bounhalli, S. Reynaud, and O. Parriaux, Optics Express **19**, 9035 (2011).
[15] M. Huang, F. L. Zhao, Y. Cheng, N. S. Xu, and Z. Z. Xu, ACS Nano **3**, 4062 (2009).
[16] J. Bonse and J. Kruger, Journal of Applied Physics **108**, 034903 (2010).
[17] G. D. Tsibidis, E. Skoulas, A. Papadopoulos, and E. Stratakis, Physical Review B **94**, 081305(R) (2016).
[18] Y. Shimotsuma, P. G. Kazansky, J. R. Qiu, and K. Hirao, Physical Review Letters **91**, 247405 (2003).
[19] B. N. Chichkov, C. Momma, S. Nolte, F. vonAlvensleben, and A. Tunnermann, Applied Physics a-Materials Science & Processing **63**, 109 (1996).
[20] B. Rethfeld, M. E. Garcia, D. S. Ivanov, and S. Anisimov, Journal of Physics D: Applied Physics **50**, 193001 (2017).
[21] E. Bevillon, R. Stoian, and J. P. Colombier, Journal of Physics-Condensed Matter **30**, 385401 (2018).
[22] G. D. Tsibidis, L. Mouchliadis, M. Pedio, and E. Stratakis, Physical Review B **101**, 075207 (2020).
[23] R. Böhme, S. Pissadakis, D. Ruthe, and K. Zimmer, Applied Physics A-Materials Science & Processing **85**, 75 (2006).
[24] Y. Jee, M. F. Becker, and R. M. Walser, Journal of the Optical Society of America B-Optical Physics **5**, 648 (1988).
[25] J. Bonse, S. Baudach, J. Kruger, W. Kautek, and M. Lenzner, Applied Physics a-Materials Science & Processing **74**, 19 (2002).
[26] S. Baudach, J. Bonse, J. Kruger, and W. Kautek, Applied Surface Science **154**, 555 (2000).
[27] S. H. Kim, I. B. Sohn, and S. Jeong, Applied Surface Science **255**, 9717 (2009).
[28] G. D. Tsibidis and E. Stratakis, Sci Rep-Uk **10**, 8675 (2020).
[29] I. M. Burakov, N. M. Bulgakova, R. Stoian, A. Rosenfeld, and I. V. Hertel, Applied Physics a-Materials Science & Processing **81**, 1639 (2005).
[30] B. Chimier, O. Utéza, N. Sanner, M. Sentis, T. Itina, P. Lassonde, F. Légaré, F. Vidal, and J. C. Kieffer, Physical Review B **84**, 094104 (2011).
[31] S. I. Anisimov, B. L. Kapeliovich, and T. L. Perel'man, Zhurnal Eksperimentalnoi Teor. Fiz. **66**, 776 (1974 [Sov. Phys. Tech. Phys. 11, 945 (1967)]).
[32] D. S. Ivanov and L. V. Zhigilei, Physical Review B **68**, 064114 (2003).
[33] R. Shwetharani, H. R. Chandan, M. Sakar, G. R. Balakrishna, K. R. Reddy, and A. V. Raghu, Int J Hydrogen Energ **45**, 18289 (2020).
[34] A. Piegari and F. o. Flory, *Optical thin films and coatings : from materials to applications*, Woodhead publishing series in electronic and optical materials, number 49.
[35] K. V. Sreekanth, S. Sreejith, S. Han, A. Mishra, X. X. Chen, H. D. Sun, C. T. Lim, and R. Singh, Nat Commun **9**, 369 (2018).

[36] M. Kumar, M. A. Khan, and H. A. Arafat, Acs Omega **5**, 3792 (2020).
[37] H. Kim, J. Proell, R. Kohler, W. Pfleging, and A. Pique, Journal of Laser Micro Nanoengineering **7**, 320 (2012).
[38] J. Proll, R. Kohler, A. Mangang, S. Ulrich, M. Bruns, H. J. Seifert, and W. Pfleging, Applied Surface Science **258**, 5146 (2012).
[39] J. Proell, R. Kohler, A. Mangang, S. Ulrich, C. Ziebert, and W. Pfleging, Journal of Laser Micro Nanoengineering **7**, 97 (2012).
[40] W. Pfleging, R. Kohler, M. Torge, V. Trouillet, F. Danneil, and M. Stuber, Applied Surface Science **257**, 7907 (2011).
[41] R. Kohler, H. Besser, M. Hagen, J. Ye, C. Ziebert, S. Ulrich, J. Proell, and W. Pfleging, Microsystem Technologies-Micro-and Nanosystems-Information Storage and Processing Systems **17**, 225 (2011).
[42] R. Kohler, P. Smyrek, S. Ulrich, M. Bruns, V. Trouillet, and W. Pfleging, Journal of Optoelectronics and Advanced Materials **12**, 547 (2010).
[43] E. Skoulas, A. C. Tasolamprou, G. Kenanakis, and E. Stratakis, Applied Surface Science **541** (2021).
[44] S. S. Wellershoff, J. Hohlfeld, J. Güdde, and E. Matthias, Applied Physics A **69**, S99 (1999).
[45] J. Hohlfeld, J. G. Muller, S. S. Wellershoff, and E. Matthias, Applied Physics B-Lasers and Optics **64**, 387 (1997).
[46] J. Güdde, J. Hohlfeld, J. G. Müller, and E. Matthias, Applied Surface Science **127**, 40 (1998).
[47] J. Hohlfeld, S. S. Wellershoff, J. Güdde, U. Conrad, V. Jahnke, and E. Matthias, Chemical Physics **251**, 237 (2000).
[48] B. C. Stuart, M. D. Feit, S. Herman, A. M. Rubenchik, B. W. Shore, and M. D. Perry, J. Opt. Soc. Am. B **13**, 459 (1996).
[49] M. Bonn, D. N. Denzler, S. Funk, M. Wolf, S. S. Wellershoff, and J. Hohlfeld, Physical Review B **61**, 1101 (2000).
[50] G. Tsibidis, D. Mansour, and E. Stratakis, *Under review* (https://arxiv.org/abs/2205.05342) (2022).
[51] A. M. Chen, L. Z. Sui, Y. Shi, Y. F. Jiang, D. P. Yang, H. Liu, M. X. Jin, and D. J. Ding, Thin Solid Films **529**, 209 (2013).
[52] A. M. Chen, H. F. Xu, Y. F. Jiang, L. Z. Sui, D. J. Ding, H. Liu, and M. X. Jin, Applied Surface Science **257**, 1678 (2010).
[53] S. Petrovic, G. D. Tsibidis, A. Kovacevic, N. Bozinovic, D. Perusko, A. Mimidis, A. Manousaki, and E. Stratakis, Eur Phys J D **75**, 304 (2021).
[54] B. Gaković, G. D. Tsibidis, E. Skoulas, S. M. Petrović, B. Vasić, and E. Stratakis, Journal of Applied Physics **122**, 223106 (2017).
[55] G. D. Tsibidis, Applied Physics Letters **104**, 051603 (2014).
[56] O. V. Kuznietsov, G. D. Tsibidis, A. V. Demchishin, A. A. Demchishin, V. Babizhetskyy, I. Saldan, S. Bellucci, and I. Gnilitskyi, Nanomaterials-Basel **11**, 316 (2021).
[57] M. Born and E. Wolf, *Principles of optics : electromagnetic theory of propagation, interference and diffraction of light* (Cambridge University Press, Cambridge ; New York, 1999), 7th expanded edn.
[58] M. A. Green, Sol Energ Mat Sol C **92**, 1305 (2008).
[59] A. D. Rakic, A. B. Djurisic, J. M. Elazar, and M. L. Majewski, Applied Optics **37**, 5271 (1998).
[60] G. D. Tsibidis, A. Mimidis, E. Skoulas, S. V. Kirner, J. Krüger, J. Bonse, and E. Stratakis, Applied Physics A **124**, 27 (2017).
[61] E. Petrakakis, G. D. Tsibidis, and E. Stratakis, Physical Review B **99**, 195201 (2019).
[62] Z. Lin, L. V. Zhigilei, and V. Celli, Physical Review B **77**, 075133 (2008).
[63] I. Gnilitskyi, T. J. Y. Derrien, Y. Levy, N. M. Bulgakova, T. Mocek, and L. Orazi, Sci Rep-Uk **7**, 8485 (2017).